\documentclass[footinbib,amsmath,amssymb,superscriptaddress,prl,reprint]{revtex4-1}

\usepackage[utf8]{inputenc}
\usepackage{graphicx}
\usepackage{xcolor}
\usepackage{hyperref}
\usepackage{notes2bib}
\usepackage{siunitx}
\DeclareSIUnit\Molar{\textsc{m}}

\begin{document}

\title{Barrier crossing in a viscoelastic bath}
\author{F\'elix Ginot}
\email{felix.ginot.uni-konstanz.de}
\affiliation{Fachbereich Physik, Universit\"{a}t Konstanz, 78457 Konstanz, Germany}
\author{Juliana Caspers}
\affiliation{Institute for Theoretical Physics, Georg-August Universit\"{a}t G\"{o}ttingen, 37073 G\"{o}ttingen, Germany}
\author{Matthias Krüger}
\affiliation{Institute for Theoretical Physics, Georg-August Universit\"{a}t G\"{o}ttingen, 37073 G\"{o}ttingen, Germany}
\author{Clemens Bechinger}
\affiliation{Fachbereich Physik, Universit\"{a}t Konstanz, 78457 Konstanz, Germany}

\begin{abstract}
We investigate the hopping dynamics of a colloidal particle across a potential barrier and within a viscoelastic, i.e., non-Markovian bath, and report two clearly separated time scales in the corresponding waiting time distributions. 
While the longer time scale exponentially depends on the barrier height, the shorter one is similar to the relaxation time of the fluid.
This short time scale is a signature of the storage and release of elastic energy inside the bath, that strongly increases the hopping rate.
Our results are in excellent agreement with numerical simulations of a simple Maxwell model.
\end{abstract}

\maketitle

The activated, i.e., fluctuation-assisted hopping of a Brownian particle across an energy barrier $\Delta U$ is one of the most important processes dominating the dynamics in many natural systems. The celebrated Kramers theory predicts the hopping rate $\nu\propto \exp(-\Delta U/k_\text{B}T)$ with $k_\text{B}T$ the thermal energy \cite{Kramers1940}. This result is in excellent agreement with activation rates measured e.g. during chemical reactions \cite{Kramers1940,Garcia-Muller2008,Fleming1986}, protein folding \cite{Laleman2017,Neupane2016,zheng2015reduction,socci1996diffusive,Satija2019}, nanomagnetic domain reversal \cite{Koch2000}, or even drug absorption \cite{Bernetti2019}. As an important extension of the Kramers' problem, several studies considered a non-Markovian bath where memory effects influence the barrier crossing dynamics ~\cite{Rondin2017,GroteHynes1980,Hanggi1982,CarmeliNitzan1984,Lavacchi2020,Kappler2019,Medina2018,Kappler2018,Ianconescu2015,Pollak1989,Goychuk2009}. Such fluids are often found in biological environments but are also widespread in technical systems.
When the bath is fully equilibrated prior to each single barrier crossing event, non-Markovian rate theory (NMRT)  predicts a Kramers-like behavior, i.e. an exponential dependence of the hopping rate on the barrier height, but with $\nu$ being considerably larger compared to an estimate based on the zero frequency memory kernel \cite{GroteHynes1980,Hanggi1982,CarmeliNitzan1984}. Indeed, such behavior has been confirmed by simulations \cite{Goychuk2009} and recent experiments \cite{Ferrer2020}. However, considering that a hopping particle leads to permanent excitation, the assumption of an equilibrated bath is not necessarily valid.

In this work we experimentally investigate the barrier crossing dynamics of a Brownian particle in an external double-well potential suspended in a viscoelastic solvent which exhibits non-Markovian behavior. For potential barriers up to several $k_\text{B}T$ we find the hopping dynamics to be characterized by {\it two} time scales which can differ by more than two orders of magnitude. While the long time scale increases exponentially with $\Delta U$ in agreement with NMRT, the short one is almost unaffected by the barrier height. The latter results from elastic energy fluctuations of the viscoelastic bath due to excitations arising from the particle's hopping motion. Our results are in quantitative agreement with a simple model where the fluid is described as a Maxwell medium.


\begin{figure}
    \centering
    \includegraphics{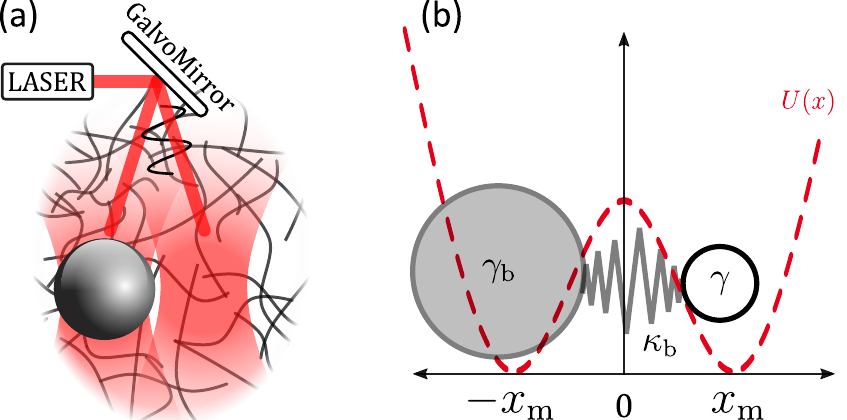}
    \caption{
    \textbf{(a)} Sketch of the experimental setup. A tightly focused laser beam is periodically deflected by a scanning mirror. It exerts a double-well potential to a $\SI{2.73}{\micro\m}$ silica tracer particle suspended in a viscoelastic solution of CPyCl-NaSal.
    \textbf{(b)} Sketch of the numerical model. A tracer particle with friction coefficient $\gamma$ is subjected to a double-well potential $U(x)$. To account for the influence of the viscoelastic bath, the tracer is coupled with linear spring of stiffness $\kappa_\text{b}$ to a bath particle with friction coefficient $\gamma_\text{b}$.
    }
    \label{fig:F1}
\end{figure}

\begin{figure}
    \centering
    \includegraphics{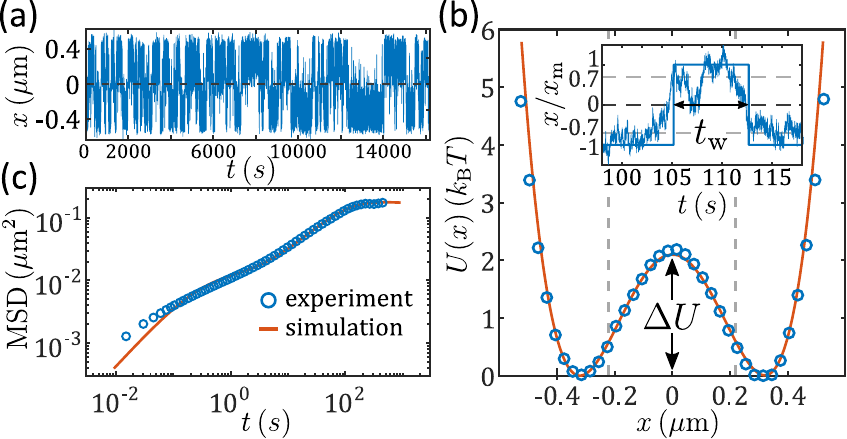}
    \caption{
    \textbf{(a)} Typical trajectory $x(t)$ of a colloidal particle in a viscoleastic fluid and subjected to a double-well potential ($\Delta U=2.1\,k_\text{B}T$) for experiments.
    \textbf{(b)} Corresponding potential obtained from experimental data (blue circles) and
    its fit used in simulations (red line). \textbf{Inset}: The particle must consecutively pass the thresholds at $\pm x_\text{c}=\pm 0.7 x_\text{m}$ to be considered as a crossing event with the waiting time  $t_\text{w}$. \textbf{(c)} Mean square displacement (MSD) of the colloidal particle obtained from experiments (blue circles) and simulations (red line). The model parameters ($\gamma$, $\gamma_\text{b}$, and $\kappa_\text{b}$) have been adjusted to yield largest agreement with the experiments.
    }
    \label{fig:F2}
\end{figure}

We perform our experiments using silica particles with diameter $\SI{2.73}{\micro\m}$ suspended in a \SI{100}{\micro\m} thick sample cell, containing a viscoelastic fluid of an \SI{5}{\milli\Molar} equimolar aqueous solution of cetylpyridinium chloride monohydrate (CPyCl) and sodium salicylate (NaSal). 
The entire sample cell was kept at \SI{25}{\celsius} during experiments.
 Under such conditions, it forms an entangled network of giant worm-like micelles leading to viscoelastic behavior with a relaxation time $ \sim \SI{0.7}{\s}$ and a zero-shear viscosity of $\eta\approx\SI{15}{\milli\pascal\s}$ determined by micro rheology measurements~\cite{Gomez-Solano2015-qu,jain2021two}.
The particle is optically trapped using a \SI{1064}{\nano \m} laser beam which is periodically deflected by a galvanostatically-driven mirror at a frequency of \SI{400}{\hertz} being focused into the sample cell with a $100\times$ high NA objective (see Fig.~\ref{fig:F1}a). To yield double-well potentials with adjustable barrier, asymmetry and well distance, the intensity of the laser beam is modulated with a phase-locked acousto-optic modulator. 
The shape of the potential $U$ and in particular the barrier height $\Delta U$ are measured from the particle's probability distribution in equilibrium (see Fig.~\ref{fig:F2}b). To avoid interactions with the surface, the optical potential was positioned in the mid-plane of the sample cell.
Particle trajectories with spatial resolution of $\sim \SI{5}{\nano\m}$ were obtained from imaged video pictures taken with a frame rate of \SI{200}{\hertz} and using a custom Matlab algorithm \cite{Crocker1994}. Each experimental run is 3-$\SI{5}{\hour}$ long, and for every potential barrier height value, we accumulate several experiments for a typical full duration of $\sim \SI{15}{\hour}$.


Before discussing our experimental results, we briefly introduce our theoretical model which provides a quantitative understanding of the experimental results.
A point-like tracer with friction coefficient $\gamma$ which mimics the colloidal particle is subjected to a double-well potential  
\begin{align}
U(x) = - 2 \Delta U \left(\frac{x}{x_\text{m}} \right)^2 + \Delta U \left(\frac{x}{x_\text{m}} \right)^4\label{eq:U}
\end{align}
with $\pm x_\text{m}$ being the positions of the potential minima and barrier height $\Delta U$ (see Fig.~\ref{fig:F1}b). 
The coupling of the tracer to the viscoelastic bath is realized by introducing another point-like particle, called bath particle (friction coefficient $\gamma_\text{b}$), which is connected to the tracer via a harmonic spring with stiffness $\kappa_\text{b}$ (see Fig.~\ref{fig:F1}b).
In the absence of any external potential, this corresponds to the well-known Maxwell-model. 
The corresponding Langevin equations for the positions $x$ and $x_\text{b}$ of tracer and bath particles thus read
\begin{eqnarray}
	\gamma\dot{x}(t) &=& -\kappa_\text{b} (x - x_\text{b}) -\nabla U  + \xi(t)  \label{LangevinEquationTracer} \\
	\gamma_{\rm b}\dot{x}_\text{b}(t) &=& -\kappa_\text{b} (x_\text{b} - x)  + \xi_{\text{b}}(t) \label{LangevinEquationBath}
\end{eqnarray}
where $\xi$ and $\xi_\text{b}$ are delta correlated random forces with zero mean ($(\xi_i,\xi_j)\in \{\xi,\xi_\mathrm{b}\}$), 
\begin{equation}
	\left\langle \xi_{i}(t) \right\rangle = 0 \quad\text{, }\quad \left\langle \xi_{i}(t)\xi_{j}(t') \right\rangle =\delta_{ij} 2k_{\rm B}T\gamma_{i}\delta(t-t')  .  \label{NoiseCorrelation}
\end{equation}
Note that the above coupled equations can be combined into one non-Markovian Langevin equation for $x$ with memory kernel $\Gamma(t)=2 \gamma \delta(t)+ \kappa_\text{b} \exp \left(- \frac{\kappa_\text{b}}{\gamma_\text{b}} t \right)$. 
Since we are also interested in the explicit position of the bath particle which contains additional important information, however, in the following we will explicitly solve the set of Eqs.~\eqref{LangevinEquationTracer} and \eqref{LangevinEquationBath}. 


Fig.~\ref{fig:F2}a shows an experimentally measured colloidal tracer trajectory $x(t)$ in a double-well potential with barrier $\Delta U=2.1\, k_\text{B}T$ and minima positions $\pm x_\text{m}=\pm$\SI{0.32}{\micro m}. During the plotted time interval of $\sim\SI{4.5}{\hour}$, approximately 200 particle jumps between the potential minima are observed. 
From the trajectories and the corresponding positional probability distribution we can compute both the external potential $U(x)$ (Fig.~\ref{fig:F2}b) and the mean square displacement (MSD, Fig.~\ref{fig:F2}c). We then compare the experimental data with the results of Eqs.~\eqref{LangevinEquationTracer},\eqref{LangevinEquationBath}, and obtain the parameters  $\gamma = \SI{0.189}{\micro \newton \s / \m}$, $\gamma_\text{b} = \SI{1.44}{\micro \newton \s / \m }$ and $\kappa_\text{b} = \SI{0.4}{\micro \newton / \m}$ which lead to the relaxation times of the tracer and the bath particle $\tau=\frac{\gamma}{\kappa_\text{b}}=\SI{0.5}{\s}$ and $\tau_\text{b} = \frac{\gamma_\text{b}}{\kappa_\text{b}}=\SI{3.6}{\s}$, respectively. We have confirmed with our experiments that these parameters do not depend on the barrier height $\Delta U$.


To analyze the hopping dynamics of the tracer particle, we have calculated the probability distribution $P(t_\text{w})$ of waiting times $t_\text{w}$, i.e. the time between two consecutive barrier crossing events. A barrier crossing event is defined by the colloid successively passing two thresholds at positions $\pm x_\text{c}=\pm 0.7 x_\text{m}$.
In Fig.~\ref{fig:F2}b (inset) we show a set of two crossing events (at $t \sim \SI{105}{s}$ and $t \sim \SI{112}{s}$), and highlight the time $t_\text{w}$ when the particle stays in the well of right.
Because the choice of $x_\text{c}$ obviously affects $P(t_\text{w})$, identical thresholds have been applied when analyzing experimental and numerical data.


\begin{figure}
    \centering
    \includegraphics{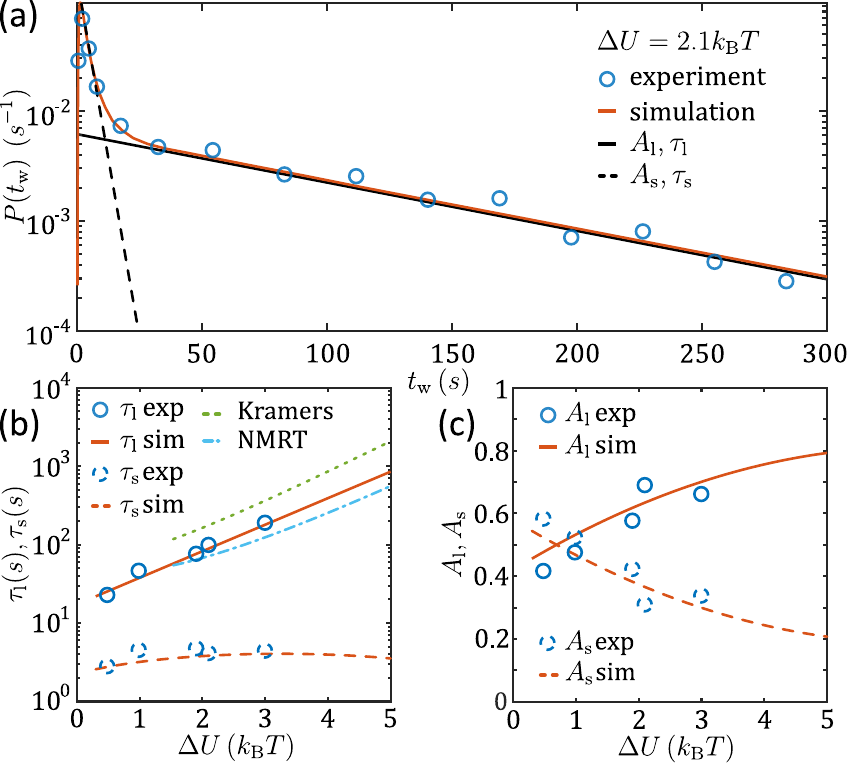}
    \caption{
    \textbf{(a)} Waiting time distribution $P(t_\mathrm{w})$ in a double-well potential ($\Delta U = 2.1 \, k_\text{B} T$) for experiments (blue circles) and simulations (red line). Both show a double-exponential decay with time scales $\tau_\mathrm{s}$ and $\tau_\mathrm{l}$, and amplitudes $A_\text{s}$ and $A_\text{l}$, respectively. Dashed and full black lines highlight the short and long time scales.
    \textbf{(b)} Short $\tau_\mathrm{s}$ (dashed lines) and long time scales $\tau_\mathrm{l}$ (full lines) for experiments (blue circles) and simulations (red lines) in double-well potentials with varying barrier height $\Delta U$. Green dotted and light blue dash-dotted lines correspond to theoretical predictions according to Kramers' and NMRT.
    \textbf{(c)} Amplitudes $A_{\rm s/l}$ of the short (dashed lines) and long (full lines) time scales $\tau_\mathrm{s}$ and $\tau_\mathrm{l}$ for different barrier heights $\Delta U$. Blue circles correspond to experiments, red lines to simulations. 
    }
    \label{fig:F3}
\end{figure}

Fig.~\ref{fig:F3}a shows $P(t_\mathrm{w})$ for a barrier height of $\Delta U=2.1\, k_\text{B} T$. Experiments (blue symbols) and simulations (red solid line) are well described by a superposition of two exponential decays, with two distinct time scales which we denote $\tau_\text{s}$ (dashed black line) and $\tau_\text{l}$ (full black line). 
\begin{equation}\label{eq:P}
    P(t_\mathrm{w}) = \frac{A_\text{s}}{\tau_\mathrm{s}} \exp \left( -\frac{t_\mathrm{w}}{\tau_\mathrm{s}} \right) + \frac{A_\text{l}}{\tau_\mathrm{l}} \exp \left( -\frac{t_\mathrm{w}}{\tau_\mathrm{l}} \right).
\end{equation}
To yield a normalized probability distribution, we set $A_\text{s}+A_\text{l}=\int_0^\infty \text{d}t P(t)\equiv 1$.
The corresponding decay times and amplitudes obtained from fitting Eq.~\eqref{eq:P} to the experimental and numerical data are shown in Figs.~\ref{fig:F3}b,c. 
Opposed to the short decay time $\tau_\mathrm{s}$ which only slightly varies with the barrier height, an exponential dependence on $\Delta U$ (qualitatively similar to Kramers' prediction, green dotted line) is observed for the longer decay time $\tau_\mathrm{l}$.
Note the good agreement with the result obtained from NMRT which is plotted as dash-dotted line in Fig.~\ref{fig:F3}b.
By definition, the amplitudes of the fast and slow decay exhibit an opposite dependence, with $A_\text{l}$ monotonically increasing with barrier height.

For a qualitative understanding of the two time scales we remind that the friction coefficient of the bath particle is much larger than that of the tracer. Therefore, the bath particle can not immediately follow a tracer's jump across the barrier. This leads to an additional elastic force on the tracer \emph{immediately} after a barrier crossing event (e.g. from left to right) which lowers the effective potential barrier. As a result, the probability for an \emph{immediate} jump back (to the left well) increases which explains the short waiting times in our experiments. If the tracer misses such an opportunity but remains longer in the (right) well than  $\tau_\text{b}$, the fluid fully relaxes, and the elastic force vanishes which again increases the effective potential barrier thus leading to long waiting times. A tracer barrier hopping event under such conditions is comparable to that in an equilibrated viscous system which also explains why $\tau_\text{l}$ exhibits a Kramers-like dependence on the potential barrier. As a consequence, the fraction of fast hopping events $A_\text{s}$ will increase with decreasing $\Delta U$ where the tracer's hopping dynamics becomes increasingly faster than $\tau_\text{b}$ in agreement with Fig.~\ref{fig:F3}c. When hopping events are rare ($\Delta U\gg k_\text{B}T$) the bath particle is likely to equilibrate prior to another tracer's barrier crossing event. Accordingly, the  amplitude of the short time scale decays to zero (Fig.~\ref{fig:F3}c).


\begin{figure}
    \centering
    \includegraphics{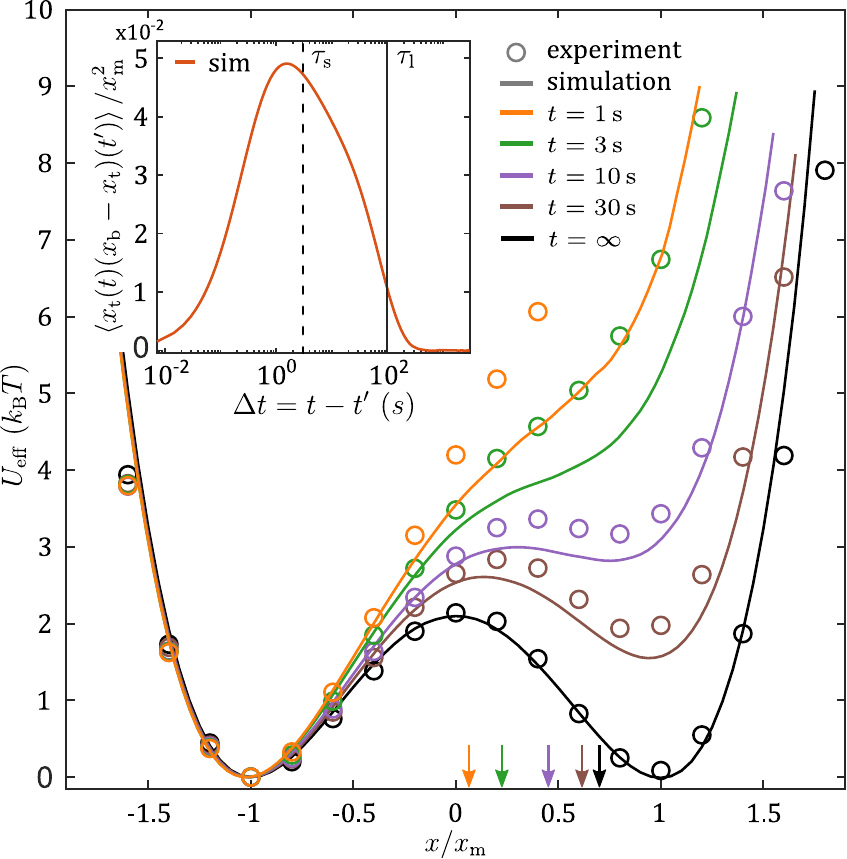}
    \caption{
    Effective pseudo potentials $U_\mathrm{eff}(x,t)$ for experiments (circles) and simulations (lines) in a double-well potential ($\Delta U = 2.1 \, k_\text{B} T$). For $t \to \infty$, $U_\mathrm{eff}(x,t)$ (colours) relaxes to $U$ (black). The arrows highlight the average bath particle position $\left< x_\text{b} \right>$ in the simulation given the tracer sits at $x=0.7 x_\text{m}$ with (colours) or without (black) conditioning on being in the left well $t$ seconds earlier. This illustrates the typical loading of the spring in the tracer-bath particle model. 
    \textbf{Inset}: Cross-correlation $\left < x(t) (x_\text{b}-x)(t') \right >$ between tracer position $x$ and bath-tracer distance $x_\text{b}-x$, highlighting the connection between the tracer position and the elastic energy stored in the system. Cross-correlation is maximum for $\Delta t = t-t' \sim \tau_\text{s}$ and decays to zero for $\Delta t > \tau_\text{l}$.
    }
    \label{fig:F4}
\end{figure}

To provide direct evidence for a time-dependent effective potential acting on the colloidal particle during barrier crossing, in a first step we have calculated the time-independent total force $F$ acting on it. It is obtained from its short-time drift motion \cite{Risken}
\begin{align}\label{eq:Feq}
    F(x)=\gamma_\infty \lim_{\delta t\to0}\int \text{d} l \frac{l}{\delta t} P(x+l, \delta t|x, 0),
\end{align}
where $\gamma_\infty$ is the colloid's short time friction coefficient and $P(x, t|x', t')$ is the two point conditional probability to find the particle at position $x$ at time $t$ given it was at position $x'$ at $t'$. The corresponding effective potential is then given by $U(x)=-\int^x F(x')\text{d}x'$. Experimentally, the value of $\gamma_\infty$ is obtained by comparing the potential $U(x)$ using integration of Eq.~\eqref{eq:Feq} with that obtained from the particle's equilibrium probability distribution (Fig.~\ref{fig:F2}b).  For the numerical simulations obviously $\gamma_\infty=\gamma$.
To consider now memory effects in the effective force acting on the particle, we replace the two point probability in Eq.~\eqref{eq:Feq} by the three point conditional probability  $P(x+l, \delta t|x, 0;x_{-t},-t)$ which adds the additional constraint that the particle was located at position $x_{-t}$ at a time interval $t$ \emph{prior} to $t=0$ where the particle was at position $x$. As a result, the drift force depends also on the time interval $t$. The difference between $F(x,-t) \equiv F(x,t)$ and $F(x)$ is then a direct measure of the non-Markovianity of the viscoelastic bath.

Fig.~\ref{fig:F4} shows the corresponding effective pseudo potential $U_\mathrm{eff}(x,t)= - \int^x F(x',t)\text{d}x'$ for the specific initial condition that the tracer particle was located a time interval $t$ \emph{prior} to $t=0$ in the left potential well, i.e. $x_{-t}\leq -x_{\text{c}}$. In agreement with our above qualitative explanation, the corresponding potentials become increasingly symmetric with increasing $t$, where the elastic coupling between the tracer and the bath particle decreases. As an immediate consequence, the mean distance between these particles $\left < x_\text{b}(t) \right >-x$ should decrease with increasing $t$. As an example we show as vertical arrows in Fig~\ref{fig:F4} how $\left < x_\text{b}(t) \right >$ gradually approaches $x=0.7 x_\text{m}$ with increasing $t$. The elastic energy corresponding to the mean tracer-bath particle distance, $U_\text{el}=1/2\kappa_\text{b}(\left < x_\text{b}(t) \right >-0.7 x_\text{m})^2$ after $t=\SI{1}{\s}$ yields $2.1\,k_\text{B}T$ (see orange arrow in Fig.~\ref{fig:F4}) which also explains the disappearance of the barrier in the effective potential. 

Interestingly, even for $t=\SI{30}{s}$ which is considerably larger than the relaxation times of the bath and tracer particles, respectively, a pronounced asymmetry in the effective potential is observed. 
This is understood by considering the cross-correlation function $\left < x(t)  (x_\text{b}-x)(t') \right >$ between tracer position $x_\text{t}$ and bath-tracer distance $x_\text{b}-x$ which is a measure of the decay of elastic correlations in the system. As seen by the inset of Fig.~\ref{fig:F4}, it only decays after more than $t=\SI{100}{s}$, corresponding to the longest time scale in our system $\tau_\text{l}$ and thus explains why the asymmetry in the effective pseudo potentials decays rather slowly.


\begin{figure}
    \centering
    \includegraphics{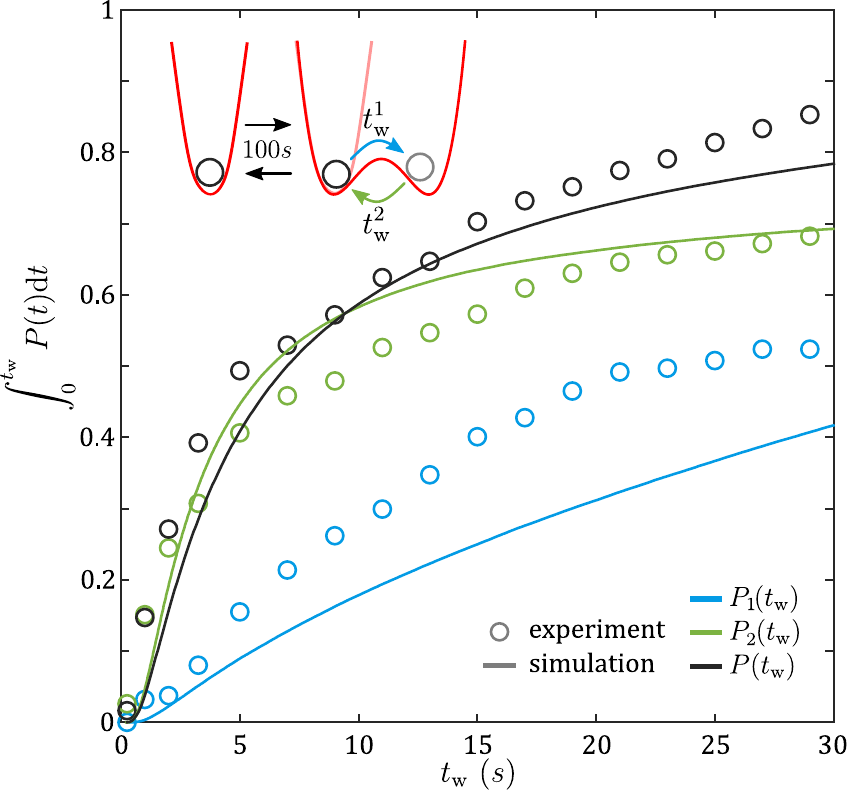}
    \caption{
    Integrated waiting time distributions $P_1(t)$ and $P_2(t)$ for experiments (circles) and simulations (lines).
    We start by forcing particle equilibration in the left well. After $\SI{100}{\s}$ we reverse the system back to its original double-well state (see sketch inset). 
    Black curve and symbols correspond to the previous situation of a static double-well potential (such as in Fig.~\ref{fig:F2}b).
    Comparing first jumps (light blue) with the static case (black), 
    We see that forcing equilibration affects the waiting times, and strongly decreases the fraction of short time jumps.
    Meanwhile, the second jumps (green), back to the first well show a behavior very similar to the static case (black).
    }
    \label{fig:F5}
\end{figure}

From the above, it follows that when a hopping process starts from a fully equilibrated bath, the short time scale should disappear, as no elastic energy is stored in the system. 
For an experimental realization of such conditions we have strongly suppressed its motion over \SI{100}{\s} which largely exceeds the system's relaxation time. This was achieved by removal of one potential well leading to a single harmonic particle trap (see inset Fig.~5). 
Having established an equilibrium condition at $t=0$, we suddenly restored the symmetric double-well potential and measured the waiting times of the first  barrier crossing. Fig.~\ref{fig:F5} shows the integrated waiting time distributions $\int^{t_\text{w}}_0 P_1(t)\text{d}t$ of first particle crossings,  which corresponds to the fraction of jumps occurring for $0 < t \leq t_\text{w}$. The data have been obtained from about 200 repetitions of the above protocol. Compared to the integrated waiting time distributions obtained in presence of a permanent double-well potential (black data), short waiting times are strongly suppressed under the constraint of an equilibrium initial state which thus confirms that the asymmetry of $U_\mathrm{eff}$ results from the elastic energy stored in the bath. Interestingly, when considering the waiting times of the second crossing ($P_2$), corresponding to jumps back to the initial well, fast hopping events are almost completely restored (green data in Fig.~\ref{fig:F5}). This demonstrates that after a single hopping event over the potential barrier the elastic energy stored in the bath has already reached its maximal value. The simulations (solid lines) reproduce this behavior, even though quantitative differences are observed (note that the experiments were performed with a different probe particle which required small adjustments in the simulation parameters). The remaining differences between experiments and simulations are most likely caused by a small shift in the position of the potential minima during the change from a single to a double-well potential.

In summary, with experiments and simulations we have investigated the hopping dynamics of an overdamped colloidal particle over an energy barrier in presence of a viscoelastic bath. Compared to a purely viscous, i.e., memory-free solvent, where the hopping dynamics is governed by the Kramers' rate, we find two time scales in the waiting time distribution for a non-Markovian bath leading to a rather fast surmounting of energy barriers as large as $3\,k_\text{B}T$. The larger time scale $\tau_\text{l}$ displays an exponential dependence on the potential barrier height $\Delta U$ in agreement with NMRT. On the opposite, the shorter time scale $\tau_\text{s}$ is caused by the permanent excitation of the bath due to the hopping motion of the tracer and is rather independent of the barrier height. Because this fast process is governed by the slow equilibration of the bath particle, $\tau_\text{s}$ should be rather independent of the specific choice of the potential shape and therefore relevant under many conditions. In particular in biological environments which are typically viscoelastic, we expect our results to be important for the correct interpretation of barrier heights derived from the particle's hopping dynamics.

\begin{acknowledgments}
We thank Matthias Fuchs for fruitful discussions and Jakob Steindl for preparing micellar solutions and performing macro-rheological measurements.
This project was funded by the Deutsche Forschungsgemeinschaft (DFG), Grant No. SFB 1432 - Project ID 425217212. F.G. acknowledges support by the Alexander von Humboldt foundation.
\end{acknowledgments}

\bibliographystyle{apsrev4-1} 

%

\end{document}